\documentclass[a4,12pt]{article}

\usepackage{amsmath,amssymb}
\usepackage{graphicx}
\usepackage{cite}

\def\tfontsize{scaled\magstep4}
\font\titlerm=cmr10 \tfontsize

\makeatletter

\@addtoreset{equation}{section}

\renewcommand{\section}{\@startsection{section}{1}{\z@}
	{-3.5ex \@plus -1ex \@minus -.2ex}{2.3ex \@plus.2ex}
	{\normalfont\normalsize\bfseries}}
\renewcommand{\subsection}{\@startsection{subsection}{2}{\z@}
	{-3.5ex \@plus -1ex \@minus -.2ex}{2.3ex \@plus.2ex}
	{\normalfont\normalsize\bfseries}}
\renewcommand{\subsubsection}{\@startsection{subsubsection}{3}{\z@}
	{-3.5ex \@plus -1ex \@minus -.2ex}{2.3ex \@plus.2ex}
	{\normalfont\normalsize\it}}
\makeatother

\newcommand\tr{\mathop{\rm tr}\nolimits}
\renewcommand\Im{\mathop{\rm Im}\nolimits}

\newcommand\vk{{\vec k}}
\newcommand\vl{{\vec l\,}}
\newcommand\vp{{\vec p\,}}
\newcommand\vq{{\vec q\,}}

\newcommand\CA{{\cal A}}
\newcommand\CN{{\cal N}}

\newcommand\CP{{\cal P}}
\newcommand\CS{{\cal S}}

\newcommand\bal{\begin{align}}
\newcommand\eal{\end{align}}

\begin{document}
\renewcommand{\thefootnote}{\fnsymbol{footnote}}

\begin{titlepage}
\begin{flushright}
\end{flushright}
\setlength{\baselineskip}{19pt}
\bigskip\bigskip\bigskip

\vbox{\centerline{\titlerm Evaluation of Entanglement Entropy}
\bigskip
\centerline{\titlerm in High Energy Elastic Scattering}
}

\bigskip\bigskip\bigskip

\centerline{Robi Peschanski\footnote{\tt robi.peschanski@ipht.fr}}
\medskip
{\it 
\centerline{ Institut de Physique Th{\' e}orique\footnote{Unit\'e Mixte de Recherche 3681 du CNRS}, Universit\'e Paris-Saclay, CEA}
\centerline{F-91191 Gif-sur-Yvette, France}
}
\bigskip
\centerline{and}
\bigskip
\centerline{Shigenori Seki\footnote{\tt sseki@mail.doshisha.ac.jp}}
\medskip
{\it 
\centerline{Faculty of Sciences and Engineering, Doshisha University}
\centerline{1-3 Tatara-Miyakodani, Kyotanabe, Kyoto 610-0394, Japan}
\medskip
\centerline{Osaka City University Advanced Mathematical Institute (OCAMI)}
\centerline{3-3-138, Sugimoto, Sumiyoshi-ku, Osaka 558-8585, Japan}
}

\vskip .3in

\centerline{\bf Abstract}
Entanglement of the  two scattered particles is expected to occur in elastic collisions, even at high energy where they are in competition with inelastic ones. We study how to evaluate quantitatively the corresponding entanglement entropy $S_{\rm EE}.$ For this sake, we regularize the divergences occurring in the formal derivation of $S_{\rm EE}$ using a regularization procedure  acting on the two-particle Hilbert space of  final states. A quantitative application is performed in proton-proton collisions at collider energies, comparing the results of $S_{\rm EE}$ with  two different cut-offs and  with a  volume-regularization obtained by a prescription fixing  the finite two-body Hilbert space volume. A significant entanglement is found which persists even at the highest available energies.

\vfill\noindent
24 June 2019; revised: 1 October 2019
\end{titlepage}

\renewcommand{\thefootnote}{\arabic{footnote}}
\setcounter{footnote}{0}
\setlength{\baselineskip}{19pt}

\section{Introduction}

Entanglement is a significant phenomenon in quantum theories  
and has been attracting many interests of scientists in various research areas. 
In this paper we are interested in the entanglement of scattering particles. 
How much are the particles entangled due to the scattering interaction? 
This is a simple and fundamental question. 
A way to answer it is to evaluate the entanglement entropy of the final state of particles. For this sake,  two-body elastic scattering appears to be a case study for entanglement in the final state.

In Ref.~\cite{PSS} the entanglement in momentum Hilbert space in the scattering process has been studied, 
and the entanglement entropy of the final state of two particles has been calculated 
in weak coupling perturbation 
by applying the method developed by Ref.~\cite{BMV} for momentum space entanglement. 
Ref.~\cite{PS} also has considered the entanglement in momentum Hilbert space 
for the elastically scattering particles, 
but has formulated non-perturbatively the entanglement entropy 
by the use of S-matrix theory \cite{Va,BV}. 
Ref.~\cite{PS}, as a result, has derived an adequate  formalism for the entanglement entropy and has suggested an   entropy formula 
of the two-particle final state after the elastic scattering. 
Additionally the entanglement entropy in this formula includes 
the influential effect of inelastic processes which are present in the overall set of the possible final states at a given high energy. 

However there is a problem of divergence in the entanglement entropy, 
which is caused by the infinite volume 
of the momentum Hilbert space in Refs,~\cite{PSS,PS}. 
Indeed the formula in Ref.~\cite{PS} is written in terms of not only physical observables, {\it i.e.}, 
the elastic and total cross sections, but also the cut-off parameter for the infinite volume. 
One of the subjects in this paper is, starting from Ref.~\cite{PS}, to formulate a finite entanglement entropy formula 
 by identifying the physical origin of the divergence in the entropy formula and using it to appropriately regularize this divergence.

As mentioned above, the entanglement in scattering process is a fundamental issue. 
For the sake of completion, we quote
some works \cite{FPAHBS,FDH,CCS,GS,KL} related to this issue, while being of a different focus than ours. 
Ref.~\cite{FPAHBS} has computed the variation of entanglement entropy 
in an elastic scattering of two interacting scalar particles at one-loop perturbative level. 
Ref.~\cite{FDH} has studied the entanglement entropy and mutual information 
in a fermion-fermion scattering.
Ref.~\cite{CCS} are concerned with quantum measurement theory and relativistic scattering theory, 
and has studied the entanglement entropy of an apparatus particle 
scattered off and a set of system particles. 
Ref.~\cite{GS} has suggested another derivation of the momentum space entanglement entropy 
in the scattering at weak coupling. 
Ref.~\cite{KL} has discussed the entanglement entropy in a deep inelastic scattering.

In our study, having performed the regularization and using the obtained  formula, it is interesting to evaluate the entanglement entropy for concrete particle scattering. We thus apply our formalism in high energy proton-proton scattering at Tevatron and  LHC energies
in order to evaluate the entanglement entropy of two-body elastic final states at the highest available energies.

The plan of our study is as follows: 
In Section 2 we reformulate the entanglement entropy of scattering particles, 
starting from Ref.~\cite{PS}, in order to determine the physical origin 
of the divergences one encounters and to properly regularize them. 
In Section 3, by using the entanglement entropy formulas obtained for different regularization procedures, 
we evaluate the regularized entanglement entropy in proton-proton scattering. 
We compare two different cut-off methods  with the case of  a volume-regularization given 
by an adequate prescription for the regularized Hilbert space volume 
without explicit cut-off procedure. 
Section 4 is devoted to a discussion of the results, an outlook on further directions and a conclusion.

\section{Formulation of entanglement entropy}

In this section, we start by recalling  the formal derivation (see Ref.~\cite{PS}) of the entanglement entropy. 
Then we reformulate the derivation in order to focus on the divergences one encounters. 
Our goal is to find the physical origin of these divergences, 
identify the divergent factor and propose the way to obtain a finite formula 
for the  entanglement entropy of the two outgoing particles.

\subsection{Density matrix}

Let us consider elastic scattering of two particles A and B 
which have { initial} 3-momentum $\vk$ and $\vl$ respectively. 
Note that in the high energy regime inelastic scattering together 
with the elastic one have a large contribution. 
In fact, both types of scattering are related through the unitarity relations.
Using the generic entanglement formalism, the statistical entanglement 
between the particles, A and B, with { final} 3-momentum respective $\vp$ and $\vq$ 
is expressed in terms of the entanglement entropy $S_{\rm EE}$ as follows: 
One starts with the  overall density matrix $\rho$ in the Hilbert space spanned 
by two-body final states 
$| \vp,\vq \rangle \equiv |\vp\rangle\!{}_A \otimes|\vq\rangle\!{}_B$.\footnote{
Although the complete relativistic quantum numbers of a particle state are denoted 
by momentum and spin (or helicity) as $|\vp,s\rangle$, 
we focus only on the momentum Hilbert space in this paper. 
We will give some comments on the helicity in Section \ref{discussconcl}.
} 
One defines a reduced density matrix as $\rho_A = \tr_B \rho ,$ where one sums over the states of particle B. 
Then the entanglement entropy is given by $S_{\rm EE} = -\tr_A \rho_A \ln
 \rho_A$, or equivalently  by $S_{\rm EE} = \lim_{n\to 1} S_{\rm RE}(n) 
 = -\lim_{n\to 1}{\partial \over \partial n}\tr_A (\rho_A)^n$, 
where $S_{\rm RE}(n) = {1 \over 1- n}\ln\tr_A(\rho_A)^n$ is the R{\' e}nyi entropy. 

The overall density matrix reads
\begin{align}
\rho &=  {1 \over \CN} \int {d^3\vp \over 2E_{A\vp}} {d^3\vq \over 2E_{B\vq}} {d^3\vp' \over 2E_{A\vp'}} {d^3\vq' \over 2E_{B\vq'}}
	\, |\vp,\vq\rangle \langle \vp,\vq|\CS|\vk,\vl\rangle \langle 
\vk,\vl|\CS^\dagger|\vp',\vq'\rangle \langle \vp',\vq'| \,,
\label{fullrho}
\end{align}
where $\CS$ is the S-matrix operator projecting the two-body initial state 
$|\vk,\vl\rangle \langle \vk,\vl|$ onto two-body final states. 
In Eq.~\eqref{fullrho}, the integration measure is the Lorentz invariant one 
$d^3\vp \over 2E_{\vp}$ for on-shell particles 
and $\CN$ is a normalization ensuring the condition $\tr_A\tr_B \rho =1$.
Tracing out $\rho$ with respect to the Hilbert space of particle B, 
we obtain the reduced density matrix, 
\begin{align}
\rho_A &\equiv \tr_B\rho = \int {d^3\vq'' \over 2E_{B\vq''}}\, {}\langle \vq'' |\rho |\vq'' \rangle \nonumber\\ 
&= {1 \over \CN} \int {d^3\vp \over 2E_{A\vp}} {d^3\vq \over 2E_{B\vq}} {d^3\vp' 
\over 2E_{A\vp'}}\, 
	\bigl(\langle \vp,\vq|\CS|\vk,\vl\rangle \langle 
\vk,\vl|\CS^\dagger|\vp',\vq\rangle \bigr) |\vp\rangle {}\langle\vp'| \,. 
\label{rhoApremier}
\end{align}
Taking into account energy-momentum conservation and the kinematics of elastic scattering 
$|\vk,\vl\rangle \to |\vp,\vq\rangle$ with $|\vk|=|\vl|=|\vp|=|\vq|$, 
one obtains
\begin{align}
\rho_A = {1 \over \CN} \int {d^3\vp \over 2E_{A\vp}}\, 
	&{\delta(0)\delta(E_{A\vp} +E_{B\vk+\vl-\vp} -E_{A\vk} -E_{B\vk}) \over 2E_{A\vp
}2E_{B\vk+\vl-\vp}} \nonumber\\ 
&\times \bigl(\langle \vp,\vk+\vl-\vp|{\bf s}|\vk,\vl \rangle \langle \vk,\vl 
|{\bf s}^\dagger|\vp,\vk+\vl-\vp\rangle \bigr) |\vp\rangle \langle\vp| \,,
\label{rhoA}
\end{align}
where $\langle \vp,\vq|\CS|\vk,\vl\rangle \equiv \delta^{(4)}(P^{(4)}_{p+q} 
-P^{(4)}_{k+l})\, \langle \vp,\vq|{\bf s}|\vk,\vl\rangle$ with the notation $P^{(4)}$ for the center-of-mass energy-momentum vector. The density matrix \eqref{rhoA} is normalized by its unit trace; 
\begin{align}
1 &= \tr_A\tr_B \rho = \tr_A \rho_A = \int {d^3\vp'' \over 2E_{A\vp''}}\, \langle \vp'' |\rho_A |\vp'' \rangle \nonumber \\
&= {1 \over \CN} \int d^3\vp\, 
	{\delta^{(4)}(0)\delta(|\vp|-|\vk|) \over 4|\vk| (E_{A\vk} +E_{B\vk})}
	\bigl|\langle \vp,-\vp |{\bf s}|\vk,-\vk \rangle \bigr|^2 \,,
\label{trace}
\end{align}
giving
\begin{align}
\CN = \delta^{(4)}(0)\, \CN' \,, \quad
\CN' = \int d^3\vp\,
	{\delta(|\vp|-|\vk|) \over 4|\vk| (E_{A\vk} +E_{B\vk})}
	\bigl|\langle \vp,-\vp |{\bf s}|\vk,-\vk \rangle \bigr|^2 \,,
\label{Nprime}
\end{align}
where Eqs.~\eqref{trace} and \eqref{Nprime} are expressed using the center-of-mass frame.
Note that the $\delta(0)$ coming from the energy 
conservation in Eq.~\eqref{rhoA} cancels the similar one in Eqs.~\eqref{Nprime},
 leaving an overall $\delta^{(3)}({0})$ due to the normalization. We shall discuss later  the potential divergence related to this 3-dimensional $\delta$-function.
 
One finally gets
\begin{align}
\rho_A = {1 \over \CN' \delta^{(3)}(0)} \int {d^3\vp \over 2E_{A\vp}}\, 
	{\delta(p-k) \over 4k (E_{A\vk} +E_{B\vk})}
	\bigl|\langle \vp,-\vp |{\bf s}|\vk,-\vk \rangle \bigr|^2
	|\vp\rangle \langle\vp| \,, 
\label{reduceddensity}
\end{align}
where 
for further purpose we quote
\begin{align}
p = |\vp|\,, \quad k = |\vk| \,, \quad  {\vp\cdot \vk \over pk} = \cos\theta \,, \label{eq:pkangle}
\end{align}
and $\theta$ is the center-of-mass scattering angle. 

\subsection{Entanglement entropy}

By performing the product of the $n$ density operators of the form \eqref{reduceddensity}, 
one obtains the formal expression for the entanglement entropy 
through the calculation of $\tr_A(\rho_A)^n$ as 
\begin{align}
\tr_A (\rho_A)^n = \int d^3\vp\, \delta^{(3)}(0)
\Biggl(\delta(p-k) 
{\bigl|\langle \vp,-\vp |{s}|\vk,-\vk \rangle 
\bigr|^2
 \over
 \CN' \delta^{(3)}(0) 4k (E_{A\vk} +E_{B\vk})}
 \Biggr)^n \,.
\label{Renyi}
\end{align}	
The overall $\delta^{(3)}(0)$ in the  integration comes from taking the trace over the A 
particle's 3-momentum.

Let us now introduce the partial wave expansion of the reduced S-matrix element \cite{Va,BV}, 
\begin{align}
\langle \vp,-\vp |{\bf s}|\vk,-\vk \rangle
&= {E_{A\vk} +E_{B\vk} \over \pi k} \cdot
	\sum_{\ell=0}^\infty (2\ell+1)(1 + 2i\tau_\ell) P_\ell (\cos \theta)\nonumber\\
&= {E_{A\vk} +E_{B\vk} \over \pi k} \cdot
	 2\biggl(\delta(1-\cos\theta) + {i \over 16\pi} \CA(s,t)\biggr) \,,
\label{matrixelementpartial}
\end{align}
where one used the known   summation formula of Legendre polynomials $P_\ell$, 
\begin{align}
\delta(1-\cos\theta) = {1 \over 2}\sum_{\ell=0}^\infty (2\ell+1) P_\ell (\cos \theta) \,, 
\label{deltafunction}
\end{align}
together with the partial wave expansion of the scattering amplitude, 
\begin{align}
{\cal A} = {16\pi} \sum_{\ell=0}^\infty (2\ell +1) \tau_\ell P_\ell (\cos \theta) \,, \label{eq:ampli}
\end{align}
and 
\begin{align}
s_\ell =1+2i\tau_\ell\,. 
\label{s-matrix}
\end{align}
is the two-body S-matrix $\ell^{\rm th}$ partial wave. 
It becomes clear from Eq.~\eqref{matrixelementpartial} 
that the powers of $\delta$-functions in Eqs.~\eqref{Nprime}, \eqref{reduceddensity} and \eqref{Renyi} give rise to divergences. 
In order to exhibit these divergences for further regularization, we 
introduce the divergent  full phase-space ``volume'', 
\begin{align}
V \equiv 2\delta(0) = \sum_{\ell=0}^\infty (2\ell +1) \,, 
\label{volume}
\end{align}
which we now prove that it is the key factor determining  all divergences 
we encounter in the derivation of the entanglement entropy.

Inserting the S-matrix element \eqref{matrixelementpartial} 
into the expression for $\CN'$ in Eqs.~\eqref{Nprime}, one obtains
\begin{align}
\CN' = {E_{A\vk} +E_{B\vk} \over \pi k} \sum_{\ell=0}^\infty (2\ell+1)|s_\ell|^2 \,.
\label{Nprimesum} 
\end{align}
With this expression one 
can reexpress Eq.~\eqref{Renyi} as
\begin{align}
\tr_A(\rho_A)^n
&= \int d^3\vp\, \delta^{(3)}(0) \Biggl[{\delta(p-k) \over \delta^{(3)}(0) 4\pi k^2} 
	{\bigl| \sum_\ell (2\ell+1)s_\ell P_\ell(\cos\theta) \bigr|^2 \over \sum_\ell (2\ell+1)|s_\ell|^2}\Biggr]^n \nonumber\\
&= \int_0^\infty dp\, 2\pi p^2 \int_{-1}^1 d\cos\theta\, {\delta(p-k) \over 4\pi k^2} 
	{\bigl| \sum_\ell (2\ell+1)s_\ell P_\ell(\cos\theta) \bigr|^2 \over \sum_\ell (2\ell+1)|s_\ell|^2} \nonumber\\
&\qquad\qquad\qquad\qquad\qquad	\times\Biggl[{\delta(p-k) \over \delta^{(3)}(0) 4\pi k^2} 
	{\bigl| \sum_\ell (2\ell+1)s_\ell P_\ell(\cos\theta) \bigr|^2 \over \sum_\ell (2\ell+1)|s_\ell|^2}\Biggr]^{n-1} \nonumber\\
&=\ \left({\delta(0) \over \delta^{(3)}(0) 2\pi k^2}\right)^{n-1}\ \int_{-1}^1 d\cos\theta\, \Biggl[{1 \over 2}
	{\bigl| \sum_\ell (2\ell+1)s_\ell P_\ell(\cos\theta) \bigr|^2 \over \sum_\ell (2\ell+1)|s_\ell|^2}\Biggr]^{n}  \,, 
\label{eq:renyiE}
\end{align}
where we reduce the momentum integration to the scattering angle 
and factorize out a constant prefactor in the last line of \eqref{eq:renyiE} between parentheses.
This prefactor can be expressed in terms of the (infinite) phase-space volume \eqref{volume},  
using the mathematical identity of $\delta$-functions 
in spherical coordinates with azimuthal symmetry,
\begin{align}
\delta^{(3)}(\vp-\vk) = {\delta(p-k) \over 4\pi k^2}\ \sum_{\ell=0}^\infty (2\ell +1) P_\ell(\cos\theta) \,. \label{delta3}
\end{align} 
In the $\cos\theta \to 1$ limit we formally obtain for the inverse prefactor in \eqref{eq:renyiE}, 
\begin{align}
2\pi k^2 \ {\delta^{(3)}(0) \over \delta(0)} = {1 \over 2}\sum_{\ell=0}^\infty (2\ell +1) = {V \over 2} \,. \label{delta2}
\end{align}

All in all we can rewrite Eq.~\eqref{Renyi} as
\begin{align}
\tr_A(\rho_A)^n &= \biggl({V \over 2}\biggr)^{1-n}\int_{-1}^1 d\cos\theta\, [\CP(\theta)]^n \,, \label{eq:trrhon} \\
\CP(\theta) &= 
	{1 \over 2}{\bigl| \sum_\ell (2\ell+1)s_\ell P_\ell(\cos\theta) \bigr|^2 \over \sum_\ell (2\ell+1)|s_\ell|^2} \,, \label{eq:Probs}
\end{align}
where, using the orthogonality property of Legendre polynomials, $\CP(\theta)$ is of norm 
\begin{align}
\int_{-1}^1 d\cos\theta\,\CP(\theta) = 1 \,.
 \label{norm}
\end{align}
Substituting Eq.~\eqref{s-matrix} into Eq.~\eqref{eq:Probs}, one writes
\begin{align}
\CP(\theta) = \delta(1-\cos\theta)\,\biggl(1 -{2 \sum_\ell(2\ell+1) |\tau_\ell|^2 \over V/2 -  \sum_\ell(2\ell+1)f_\ell}\biggr)
+{\bigl|\sum_\ell(2\ell +1)\tau_\ell P_\ell(\cos\theta)\bigr|^2 \over V/2 -\sum_\ell(2\ell+1)f_\ell} \,,
 \label{eq:Pzeta}
\end{align}
where the $f_\ell$ are the partial wave components of the inelastic cross section related to the elastic ones $\tau_\ell$ through the unitarity relation, $s_\ell s_\ell^* = 1-2f_\ell$, 
or equivalently 
\begin{align}
f_\ell= 2\bigl(\Im\tau_\ell- |\tau_\ell|^2 \bigr) \,. \label{eq:unitarityPW}
\end{align}
Indeed,  the standard expressions for physical scattering observables in terms of partial wave  components $\tau_\ell$ and $f_\ell$ read
\begin{align}
\sigma_{\rm tot}  = {4\pi \over k^2} \sum_{\ell=0}^\infty (2\ell +1) \Im \tau_\ell\,, \quad
\sigma_{\rm el} = {4\pi \over k^2} \sum_{\ell=0}^\infty (2\ell +1)|\tau_\ell|^2 \,, \quad 
\sigma_{\rm inel} = {2\pi \over k^2} \sum_{\ell=0}^\infty (2\ell +1) f_\ell 
\label{eq:crosssec}
\end{align}
and 
\begin{align}
{d\sigma_{\rm el} \over dt} = {\pi \over k^4} \biggl| \sum_{\ell}(2\ell + 1)\tau_\ell P_\ell(\cos\theta) \biggr|^2
= {|{\cal A}|^2 \over 256\pi k^4} \,, \label{eq:difcrosssec}
\end{align}
where the Mandelstam variable $t = 2k^2(\cos\theta-1)$.
We finally find the following expression for $\CP(\theta)$;
\begin{align}
\CP(\theta) = \delta(1-\cos\theta)\cdot\biggl(1-{\sigma_{\rm el} \over {\pi V / k^2} - \sigma_{\rm inel}}\biggr)
+{1 \over \sigma_{\rm el}}{d\sigma_{\rm el} \over d\cos\theta}
\cdot\biggl({\sigma_{\rm el} \over {\pi V / k^2} - \sigma_{\rm inel}}\biggr) \,.
 \label{eq:Pzetanew}
\end{align}

Using Eq.~\eqref{eq:trrhon}, we write formally the entanglement entropy as 
\begin{align}
S_{\rm EE} = - \lim_{n \to 1} {\partial \over \partial n} \tr_A (\rho_A)^n 
= \ln {V \over 2} -\int_{-1}^1 d\cos\theta\, \CP(\theta) \ln \CP(\theta) \,.
\label{entropy}
\end{align}

From Eqs.~\eqref{eq:Pzetanew} and \eqref{entropy}, we observe that   the divergences, 
in particular those due to the product of the $\delta$-functions contained 
in $[\CP(\theta)]^n$ in the definition of $\tr_A(\rho_A)^n$ in Eq.~\eqref{eq:trrhon}, 
are related to the  infinite phase-space ``volume'' $V\,(=\infty)$ defined in Eq.~\eqref{volume}. 
In this case $\CP(\theta)$ reduces to $\delta(1-\cos\theta)$  
and the entanglement entropy is zero. 
However, It is physically obvious that at each center-of-mass energy, 
only a finite (of order $const.\times k^2$) number of partial waves contribute 
to the final interacting states. 
Indeed, in the formal calculation of the entanglement entropy we performed, 
all two-body states in the Hilbert space have been included for the summation of final states, 
whether they come from the interaction or not. 
Therefore we have to restore a projection of the two-body Hilbert space onto the set of interacting ones. 
We are thus led to interpret the divergence due to $\delta$-functions 
and the ``volume'' $V$, as due to the infinite number of non-interacting two-body states. Hence an appropriate regularization is required.

\subsection{Volume-regularization}

As we pointed out in the previous subsection, 
the first term in Eq.~\eqref{eq:Pzetanew} comes from the part of the two-body Hilbert space 
of the final states which does not correspond to the interacting states at the given energy. 
In an ideal cut-off independent way to avoid such non-interacting modes, we are led to note that
the volume $V$ could be regularized to ${\tilde V}$ so that the first term vanishes, {\it i.e.}, 
\begin{align}
{\tilde V} = {k^2 \sigma_{\rm tot} \over \pi} \,, \quad
{\tilde \CP}(\theta) = {1 \over \sigma_{\rm el}}{d\sigma_{\rm el} \over d\cos\theta}
= {2k^2 \over \sigma_{\rm el}}{d\sigma_{\rm el} \over dt}\ . \label{eq:regprob}
\end{align}
We call it the volume-regularization assumption.

From the second equation in Eqs.~\eqref{eq:regprob}, 
and recalling the normalization condition \eqref{norm}, 
one realizes that  ${\tilde \CP}(\theta)$ can be interpreted 
as the physical probability of interaction.

The relations \eqref{eq:regprob} lead 
the formal entanglement entropy \eqref{entropy} 
to the  volume-regularized entanglement entropy, 
\begin{align}
{\tilde S}_{\rm EE} &= -\int_{-\infty}^0 dt\, {1 \over \sigma_{\rm el}} {d\sigma_{\rm el} \over dt} \ln \biggl({4\pi \over \sigma_{\rm tot} \sigma_{\rm el}} {d\sigma_{\rm el} \over dt}\biggr)  \,. \label{eq:EEtilde}
\end{align}
However currently we do not know yet which could be an effective regularization  of  the partial wave components leading to the volume-regularization without modifying the observables. The volume-regularization can thus be 
 called  ideal, 
since  it only depends on measurable observables, and not on  any cut-off.
In the following sections, we try some concrete regularization methods, 
in order to obtain an  approximation of the ideal determination \eqref{eq:EEtilde} of the entanglement entropy and compare it with  the one obtained from Eq.~\eqref{eq:EEtilde}.


\section{Evaluation of the regularized entanglement entropy}

\subsection{Cut-off regularization}

We shall make use of the impact parameter $b$ 
and the corresponding representation of observables, 
which correspond to a description of high-energy scattering observables, appropriate to our goal. 
The scattering amplitude \eqref{eq:ampli} by the partial wave expansion is rewritten 
in the impact-parameter representation as 
\begin{align}
\CA = 16\pi \sum_{\ell=0}^\infty (2\ell+1)\tau_\ell P_\ell(\cos\theta) 
= 32\pi k^2\int_0^\infty bdb\,\tau(b)J_0(b\sqrt{-t}) \,, \label{eq:SAimpact}
\end{align}
where $J_n$ is the well-known Bessel function of order $n$. 
In other words, $\tau(b)$ is defined by this equation.

In actual physics experiments, $\tau_\ell$ for large $\ell$, 
{\it i.e.}, $\tau(b)$ for large $b$ (because of $bk \sim \ell$),
does not contribute to the scattering amplitude. 
Therefore we are led to a regularization truncating the large $b$ modes 
by introducing a cut-off function $c(b)$ satisfying $\lim_{b\to\infty} c(b) = 0$, so that the amplitude becomes 
\begin{align}
{\hat \CA} = 32\pi k^2\int_0^\infty bdb\,c(b) \tau(b)J_0(b\sqrt{-t}) \,. \label{eq:COscatamp}
\end{align}
This prescription gives an approximation of physical Hilbert space. 
Following this scattering amplitude, 
the differential elastic cross section becomes
\begin{align}
{d{\hat \sigma}_{\rm el} \over dt} = 
{4\pi}\biggl| \int_{0}^\infty bdb\, c(b) \tau(b) J_0(b\sqrt{-t})\biggr|^2  \,, \label{eq:regdifcrosssec}
\end{align}
and the total, elastic and inelastic cross sections become
\begin{align}
{\hat \sigma}_{\rm tot} &= 8\pi \int_0^\infty bdb\, c^2(b) \Im\tau(b) \,, \label{eq:regtotcs}\\
{\hat \sigma}_{\rm el} &= \int_{-\infty}^0 dt\, {d{\hat \sigma}_{\rm el} \over dt}
	=  8\pi \int_0^\infty bdb\, c^2(b) |\tau(b)|^2 \,, \label{eq:regelcs} \\
{\hat \sigma}_{\rm inel} &= 4\pi \int_0^\infty bdb\, c^2(b) f(b) \,. \label{eq:reginelcs}
\end{align}
Since the relation ${\hat \sigma}_{\rm tot} = {\hat \sigma}_{\rm el} + {\hat \sigma}_{\rm inel}$ is preserved by the regularization, 
$f(b)$ is written in terms of $\tau(b)$ as $f(b) = 2(\Im\tau(b)-|\tau(b)|^2)$. 
This expression in the impact parameter space corresponds to Eq.~\eqref{eq:unitarityPW}. 

Under the cut-off approximation, the volume of the regularized Hilbert space ${\tilde V}$ is 
\begin{align}
{\tilde V} \approx {\hat V} = {k^2 \over \pi} {\hat \sigma}_{\rm tot} \,. \label{eq:stcond}
\end{align}
and the entanglement entropy \eqref{eq:EEtilde} is
\begin{align}
{\hat S}_{\rm EE} &= -\int_{-\infty}^0 dt\, {1 \over {\hat \sigma}_{\rm el}} {d{\hat \sigma}_{\rm el} \over dt} \ln \biggl({4\pi \over {\hat \sigma}_{\rm tot} {\hat \sigma}_{\rm el}} {d{\hat \sigma}_{\rm el} \over dt}\biggr)  \,. \label{eq:regeeco}
\end{align} 

It is important to note that ${\tilde \CP}(\theta)$ in Eq.~\eqref{eq:regprob} 
keeps to be a finite probability distribution verifying positivity 
and unit norm even under the cut-off approximation, 
{\it i.e.}, 
\begin{align}
{\hat \CP}(\theta) &= {2k^2 \over {\hat \sigma}_{\rm el}}{d{\hat \sigma}_{\rm el} \over dt} \,,
\end{align}
since Eq.~\eqref{eq:regelcs} leads to 
\begin{align}
\int_{-1}^{1}d\cos\theta\, {\hat \CP}(\theta) &= \int_{-\infty}^0dt\, {1 \over {\hat \sigma}_{\rm el}}{d{\hat \sigma}_{\rm el} \over dt} = 1 \,.
\end{align}

\subsubsection{Step-function cut-off}

In order for a concrete evaluation of the entanglement entropy, 
we, for instance, employ a step-function as the simplest cut-off function:
\begin{align}
c(b) = 
\begin{cases}
1 & (b \leq 2\Lambda) \\ 
0 & (b > 2\Lambda)
\end{cases}\,. \label{eq:stepCO}
\end{align}
The scattering amplitude \eqref{eq:COscatamp} becomes 
\begin{align}
{\hat \CA} = 32\pi k^2\int_0^{2\Lambda} bdb\,\tau(b)J_0(b\sqrt{-t}) \,. 
\end{align}
This cut-off truncates the modes whose impact parameter is larger than 
the maximal impact parameter $2\Lambda$. 
Since the impact parameter $b$ is related with angular momentum $\ell$ by $b=\ell/k$, 
$\ell$ has an upper bound $L$ defined by $2\Lambda k \equiv L$. 
Therefore one can also recognize the scattering amplitude as ${\hat \CA} = 16\pi \sum_{\ell=0}^L (2\ell+1) \tau_\ell P_\ell(\cos\theta)$. 
Simultaneously the cut-off regularizes the infinite volume $V$ of the full Hilbert space as 
\begin{align}
{\hat V} = 2k^2 \int_0^{2\Lambda} bdb = 4k^2\Lambda^2 \,.
\end{align}
Then the condition \eqref{eq:stcond} determines $\Lambda$ such that 
\begin{align}
4\pi \Lambda^2 = {\hat \sigma}_{\rm tot} \,. \label{eq:crLambda}
\end{align}
Under the cut-off \eqref{eq:stepCO} 
we write the differential cross section \eqref{eq:regdifcrosssec}, 
the total cross section \eqref{eq:regtotcs} 
and the elastic cross section \eqref{eq:regelcs} as
\begin{align}
{d{\hat \sigma}_{\rm el} \over dt} &= 
{4\pi}\biggl| \int_{0}^{2\Lambda} bdb\, \tau(b) J_0(b\sqrt{-t})\biggr|^2 \,, \label{STdifCS}\\
{\hat \sigma}_{\rm tot} &= 8\pi \int_0^{2\Lambda} bdb\, \Im\tau(b) \,, \label{STtotCS} \\
{\hat \sigma}_{\rm el} &= 8\pi \int_0^{2\Lambda} bdb\, |\tau(b)|^2 \,. \label{STelCS}
\end{align}

\subsubsection{Gaussian cut-off}

By concrete comparison with the step-function cut-off, 
let us consider a Gaussian cut-off function;
\begin{align}
c(b) = \exp{\biggl(-{\small {1 \over 2}}\cdot{b^2 \over 4\Lambda^2}\biggr)} \,, \label{eq:gaussco}
\end{align}
corresponding to an impact-parameter width $2\Lambda$.
Then the differential cross section \eqref{eq:regdifcrosssec}, 
the total cross section \eqref{eq:regtotcs} 
and the elastic cross section \eqref{eq:regelcs} become
\begin{align}
{d{\hat \sigma}_{\rm el} \over dt} &= 
{4\pi}\biggl| \int_0^\infty bdb\, e^{-{b^2 \over 8\Lambda^2}} \tau(b) J_0(b\sqrt{-t})\biggr|^2 \,, \label{GAdifCS}\\
{\hat \sigma}_{\rm tot} &= 8\pi \int_0^\infty bdb\, e^{-{b^2 \over 4\Lambda^2}} \Im\tau(b) \,, \label{GAtotCS} \\
{\hat \sigma}_{\rm el} &= 8\pi \int_0^\infty bdb\, e^{-{b^2 \over 4\Lambda^2}} |\tau(b)|^2 \,. \label{GAelCS}
\end{align}
Since \eqref{eq:stcond} shows that the Hilbert space volume is regularized 
in the same way as the total cross section, 
the regularized Hilbert space volume under the Gaussian cut-off \eqref{eq:gaussco} becomes 
\begin{align}
{\hat V} = 2k^2\int_0^\infty bdb\, c^2(b) = 4k^2\Lambda^2 \,,
\end{align}
and the condition \eqref{eq:stcond} is written as 
\begin{align}
4\pi \Lambda^2 = {\hat \sigma}_{\rm tot} \,. \label{eq:gaussLam}
\end{align}
This condition has the same expression as the one \eqref{eq:crLambda} 
in the step function cut-off.

\subsection{Application: the diffraction peak approximation in proton-proton scattering at high energy}

We concentrate on the proton-proton scattering, because we can use 
the experimental data given by the Tevatron (at $\sqrt{s} = 1800$\,GeV) and the LHC (at $\sqrt{s} = 7000$, 8000, 13000\,GeV), of which data 
are listed in Table~\ref{tab:highenedata}. 
Note that the difference between ${\bar p}$-$p$ and $p$-$p$ scattering at the Tevatron and LHC energies is not expected to be relevant in our study and thus has been neglected.
\begin{table}
\begin{center}
\begin{tabular}{|c||c|c|c|}
\hline
$\sqrt{s}$ [GeV] & $\sigma_{\rm tot}$ [mb] & $\sigma_{\rm el}$ [mb] & Refs. \\ \hline  
\phantom{1}1800 & \phantom{0}72.1\phantom{0} & 16.6\phantom{0} & \cite{Tevatroni,Tevatronii} \\ 
\phantom{1}7000 & \phantom{0}98.58 & 25.43 & \cite{TOTEMi} \\
\phantom{1}8000 & 101.7\phantom{0} & 27.1\phantom{0} & \cite{TOTEMii,TOTEMiii} \\ 
13000 & 110.6\phantom{0} & 31.0\phantom{0} & \cite{TOTEMiv} \\
\hline
\end{tabular}
\end{center}
\caption{Experimental  cross sections by Tevatron and LHC, central values.}\label{tab:highenedata}
\end{table}

Since we must know the differential cross section ${d\sigma_{\rm el} \over dt}$ as a function of $t$ 
in order to evaluate the entanglement entropy \eqref{eq:regeeco}, 
here we assume the diffraction peak model, 
which is described by the following scattering amplitude:
\begin{align}
\CA(s,t) = is \sigma_{\rm tot} e^{{1 \over 2}Bt} \,, \label{eq:scatamp}
\end{align}
where $B$ is the slope parameter. 
We assume sufficiently high energy, so that $s \approx 4k^2$. 
The differential elastic cross section is 
\begin{align}
{d\sigma_{\rm el} \over dt} = {\sigma_{\rm tot}^2 \over 16\pi} e^{Bt} \,,  \label{eq:difpeakdifCS}
\end{align}
and the elastic cross section is 
\begin{align}
\sigma_{\rm el} = \int_{-\infty}^0dt\,{d\sigma_{\rm el} \over dt} 
= {\sigma_{\rm tot}^2 \over 16\pi B} \,.
\end{align}
Therefore the slope parameter $B$ can be written in terms of $\sigma_{\rm tot}$ 
and $\sigma_{\rm el}$ as 
\begin{align}
B = {\sigma_{\rm tot}^2 \over 16\pi \sigma_{\rm el}} \,. \label{eq:difpeakslope}
\end{align}
From Eq.~\eqref{eq:SAimpact} and \eqref{eq:scatamp}, 
$\tau(b)$ is calculated, 
\begin{align}
\tau(b) = {1 \over 32\pi k^2} \int_0^\infty \sqrt{-t} d\sqrt{-t}\, \CA(s,t) J_0(b\sqrt{-t}) 
= i {\sigma_{\rm tot} \over 8\pi B} e^{-{b^2 \over 2B}}  \,. \label{eq:difpeaktau}
\end{align}

\subsubsection{Step-function cut-off}

In terms of Eq.~\eqref{eq:difpeaktau} we write down the truncated differential cross section \eqref{STdifCS}, 
\begin{align}
{d{\hat \sigma}_{\rm el} \over dt} 
= {\sigma_{\rm tot}^2 \over 16\pi B^2} \biggl( \int_0^{2\Lambda} bdb\, e^{-{b^2 \over 2B}} J_0(b\sqrt{-t}) \biggr)^2
\end{align}
and compute the truncated cross sections \eqref{STtotCS} and \eqref{STelCS}, 
\begin{align}
{\hat \sigma}_{\rm tot} = \sigma_{\rm tot} \Bigl( 1-e^{-{2 \over B}\Lambda^2} \Bigr) \,, \quad
{\hat \sigma}_{\rm el} = {\sigma_{\rm tot}^2 \over 16\pi B}\left(1-e^{-{4\over B}\Lambda^2 }\right) \,.
\label{eq:difpeakCS}
\end{align}
Then the condition \eqref{eq:crLambda} determining $\Lambda$ becomes 
\begin{align}
{4\pi \Lambda^2 \over \sigma_{\rm tot}} 
=  1-e^{-{2\over B}\Lambda^2} \,.
\end{align}
By using the data in Table~\ref{tab:highenedata}, we numerically calculate 
the cut-off parameter $\Lambda$, the truncated cross sections \eqref{eq:difpeakCS} 
and the entanglement entropy \eqref{eq:regeeco}, 
and the results are shown in Table~\ref{tab:STresult}.
\begin{table}
\begin{center}
\begin{tabular}{|c||c|c|c|c||c|}
\hline
$\sqrt{s}$ [GeV] & $\Lambda$ [fm] & ${\hat \sigma_{\rm tot}}$ [mb] & ${\hat \sigma_{\rm el}}$ [mb] & ${\hat S}_{\rm EE}$ & $B$ [GeV$^{-2}$] \\ \hline 
\phantom{1}1800 & 0.6550 & 53.91 & 15.54 & 1.193 & 16.00 \\ 
\phantom{1}7000 & 0.7988 & 80.18 & 24.54 & 1.192 & 19.52 \\
\phantom{1}8000 & 0.8192 & 84.34 & 26.31 & 1.197 & 19.50 \\ 
13000 & 0.8659 & 94.23 & 30.32 & 1.212 & 20.16 \\
\hline
\end{tabular}
\end{center}
\caption{The cut-off ($\Lambda$), the cross sections (${\hat \sigma}_{\rm tot}$, 
${\hat \sigma}_{\rm el}$) and the entanglement entropy (${\hat S}_{\rm EE}$) in the step-function regularization. The slope $B$ is calculated by Eq.\eqref{eq:difpeakslope} 
from the experimental data of $\sigma_{\rm tot}$ and $\sigma_{\rm el}$. }\label{tab:STresult}
\end{table}

\subsubsection{Gaussian cut-off}

The differential cross section \eqref{GAdifCS} truncated by the Gaussian cut-off 
with Eq.~\eqref{eq:difpeaktau} is written down as 
\begin{align}
{d{\hat \sigma}_{\rm el} \over dt} 
= {\sigma_{\rm tot}^2 \over 16\pi} \biggl( 1 + {B \over 4\Lambda^2}\biggr)^{-2} 
\exp\Biggl( {B \over 1 + {B \over 4\Lambda^2}}t \Biggr) \,. 
\end{align}
In the same way we calculate the truncated cross sections \eqref{GAtotCS} and \eqref{GAelCS}, so that 
\begin{align}
{\hat \sigma}_{\rm tot} = \sigma_{\rm tot}\biggl(1 +{B \over 2\Lambda^2} \biggr)^{-1} \,, \quad
{\hat \sigma}_{\rm el} = {\sigma_{\rm tot}^2 \over 16\pi B} \biggl(1 +{B \over 4\Lambda^2} \biggr)^{-1} \,.
\end{align}
The condition \eqref{eq:crLambda} fixes $\Lambda$ as 
\begin{align}
\Lambda = \sqrt{{\sigma_{\rm tot} \over 4\pi} -{B \over 2} } \,. 
\end{align} 
Furthermore one can write down the entanglement entropy \eqref{eq:regeeco} as 
\begin{align}
{\hat S}_{\rm EE} = 1- \ln {4\pi B \bigl(1 +{B \over 2\Lambda^2} \bigr) \over \sigma_{\rm tot} \bigl(1 +{B \over 4\Lambda^2} \bigr)} \,. 
\end{align}
The numerical evaluation of the cut-off, the total and elastic cross sections 
and the entanglement entropy are shown in Table~\ref{tab:GAresult}. 
\begin{table}
\begin{center}
\begin{tabular}{|c||c|c|c|c|}
\hline
$\sqrt{s}$ [GeV] & $\Lambda$ [fm] & ${\hat \sigma_{\rm tot}}$ [mb] & ${\hat \sigma_{\rm el}}$ [mb] & ${\hat S}_{\rm EE}$ \\ \hline  
\phantom{1}1800 & 0.5121 & 32.96 & 10.41 & 0.6009 \\ 
\phantom{1}7000 & 0.6359 & 50.81 & 17.30 & 0.7539 \\
\phantom{1}8000 & 0.6555 & 53.99 & 18.79 & 0.7965 \\ 
13000 & 0.6983 & 61.28 & 22.10 & 0.8621 \\
\hline
\end{tabular}
\end{center}
\caption{The cut-off ($\Lambda$), the cross sections (${\hat \sigma}_{\rm tot}$, 
${\hat \sigma}_{\rm el}$) and the entanglement entropy (${\hat S}_{\rm EE}$) in the Gaussian regularization.}\label{tab:GAresult}
\end{table}

\subsubsection{Comparison with  volume-regularization}

In order to compare the cut-off regularizations with the volume-regularization, 
let us try to evaluate the entanglement entropy ${\tilde S}_{\rm EE}$ 
in Eq.~\eqref{eq:EEtilde} by the volume-regularization. 
Although we do not know how to concretely realize the volume-regularization, 
we compute ${\tilde S}_{\rm EE}$ by the use of ${d\sigma_{\rm el} \over dt}$ 
given by Eq.~\eqref{eq:difpeakdifCS} with Eq.~\eqref{eq:difpeakslope} in the diffraction peak model. 
Then the entanglement entropy \eqref{eq:EEtilde} becomes 
\begin{align}
{\tilde S}_{\rm EE} = 1 +\ln{4\sigma_{\rm el} \over \sigma_{\rm tot}} \,. \label{eq:seeidealdif}
\end{align}
Evaluating this in terms of the data in Table \ref{tab:highenedata}, 
we show the results in Table \ref{tab:idealSEE}. 
\begin{table}
\begin{center}
\begin{tabular}{|c||c|}
\hline
$\sqrt{s}$ [GeV] & ${\tilde S}_{\rm EE}$ \\ \hline  
\phantom{1}1800 & 0.9176 \\ 
\phantom{1}7000 & 1.031\phantom{0} \\
\phantom{1}8000 & 1.063\phantom{0} \\ 
13000 & 1.114\phantom{0} \\
\hline
\end{tabular}
\end{center}
\caption{The entanglement entropy in the volume-regularization.}\label{tab:idealSEE}
\end{table}
The entanglement entropy monotonically increases according 
as the center-of-mass energy becomes higher.

\begin{figure}
\begin{center}
\includegraphics{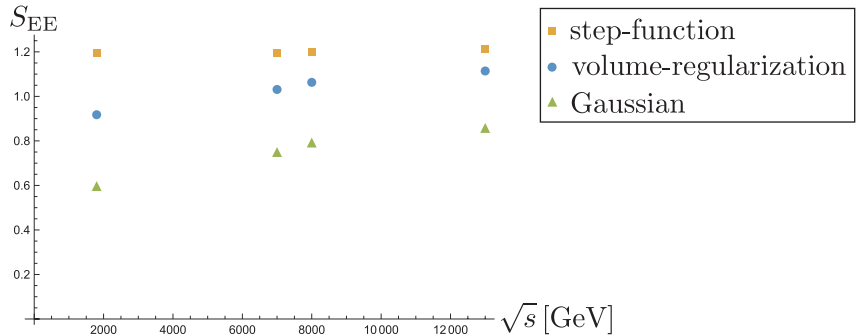}
\end{center}
\caption{The entanglement entropy in three different regularizations with respect to the center-of-mass energy.}
\label{fig:plots}
\end{figure}

The truncated cross sections in Table \ref{tab:STresult} by the step-function cut-off 
 give a closer approximation to the experimental data in Table \ref{tab:highenedata} 
better than those in Table \ref{tab:GAresult} by the Gaussian cut-off. 
As shown in Fig.~\ref{fig:plots}, actually the entanglement entropy obtained from the volume-regularization appears to be framed by  the step-function one (above) and the Gaussian one (below).

\section{Discussion, conclusion and outlook} \label{discussconcl}

In our study, we have evaluated the entanglement entropy $S_{\rm EE}$ for the two particles 
elastically produced in a high-energy collision. 
For this sake, we have used a  regularization procedure, 
in order to get rid of the divergences appearing in the formal derivation of $S_{\rm EE}$. 
These divergences happen to be  related to the infinite ``volume'' 
of the full two-particle Hilbert space, be there coming from the interaction or not. 
It can be regularized by considering the finite two-particle Hilbert space 
actually spanned by elastic collisions at a given energy. 
For the discussion we have first introduced  a formulation of a finite entanglement entropy ${\tilde S}_{\rm EE}$ 
using the formal definition supplemented with a regularized Hilbert space volume, 
which is defined by  projecting out the volume of phase space 
spanned by the non-interacting final states responsible of the divergence. 
We then considered two explicit cut-off definitions, 
one using a step-function and the other with a Gaussian. 

Summarizing our results, we found the following:
\begin{itemize}
\item[i)] 
The  volume-regularized formulation provides an expression of the entanglement entropy in terms of physical observables \eqref{eq:EEtilde};
 \begin{align}
{\tilde S}_{\rm EE} &= -\int_{-\infty}^0 dt\, {1 \over \sigma_{\rm el}} {d\sigma_{\rm el} \over dt} \ln \biggl({4\pi \over \sigma_{\rm tot} \sigma_{\rm el}} {d\sigma_{\rm el} \over dt}\biggr)  \,. \nonumber
\end{align}
\item[ii)] 
In search of an adequate quantitative cut-off procedure 
defining the finite physical Hilbert space, 
we considered the case of proton-proton elastic scattering at the Tevatron and LHC energies. 
In a diffraction peak approximation as a simple example, 
we have compared the numerical results for the regularized entanglement entropy 
${\hat S}_{\rm EE}$ in two different cut-offs, 
and we also compared them with the result 
for the entanglement entropy ${\tilde S}_{\rm EE}$ (see Eq.~\eqref{eq:seeidealdif}) 
from the  volume-regularization. 
\item[iii)] 
Since a cut-off dependence appears for the observables in the formula \eqref{eq:EEtilde} 
and modifies their contribution to the entanglement entropy, 
the effect of the cut-off is to replace the observables in Eq.~\eqref{eq:EEtilde} 
with their expressions with the cut-off as ${\hat S}_{\rm EE}$ in Eq.~\eqref{eq:regeeco}. 
The step-function cut-off appears to give a better approximation of the real observables  than the Gaussian one. However, the result for the entanglement entropy  ${\tilde S}_{\rm EE}$ boils down to a framing of the volume-regularized entropy by the step-function one (above) and the Gaussian one (below).
\item[iv)]
The trend of the overall results for ${\tilde S}_{\rm EE}$ clearly demonstrates
a non-zero entanglement entropy showing that a non-negligible entanglement 
is generated in a high-energy elastic collision, 
even in the presence of a large sector of inelastic reactions. 
Indeed, the entanglement entropy is different from zero and stays around unity, 
while increasing slightly with the center-of-mass energy. 
For instance, in the diffraction peak approximation, the volume-regularization gives Eq.~\eqref{eq:seeidealdif};
\begin{align}
{\tilde S}_{\rm EE} = 1+2\ln 2 + \ln{\biggl({\sigma_{\rm el}\over \sigma_{\rm tot} }\biggr) }\,, \nonumber
\end{align}
which allows one to relate the entanglement entropy simply to the ratio 
${\sigma_{\rm el} \over \sigma_{\rm tot}}$. 
Higher is the ratio, larger is the entanglement entropy, which seems physically sound. 
Moreover, it is known that this ratio stays experimentally around 1/4, 
and thus ${\tilde S}_{\rm EE} \sim 1$.
\end{itemize}

As an outlook, it would be useful to find a better cut-off procedure, 
which would leave the observables unchanged or only slightly changed by the regularization procedure. 
For example, an optimization calculation could be introduced to define the cut-off 
less arbitrarily as those we chose in our present study. 
Then the full set of experimental observables could be safely introduced 
in the calculation of ${\tilde S}_{\rm EE}$, 
without cut-off dependence and diffraction peak approximation.

We have considered the entanglement entropy in the momentum Hilbert space. 
However the relativistic state of a particle also have a quantum number of spin (or helicity). 
Therefore one can consider the entanglement entropy in the momentum and spin Hilbert space, 
{\it i.e.}, $\{|\vp,s\rangle\!_A\} \otimes \{|\vq,s'\rangle\!_B\}$. 
In a similar way as what we studied in this paper, 
such entropy will be formulated by the use of the S-matrix with respect to the helicity, 
which was studied in some literature \cite{GPSS,CDMP}.
Especially in high energy scattering of hadrons, where Pomeron exchange is dominant, 
the s-channel helicities of the particles are mostly conserved, 
and thus adding the spin boils down to extend our analysis to elastic scattering of
quantum states with given momentum and given s-channel helicity.

\bigbreak\bigskip\bigskip
\centerline{{\bf Acknowledgments}}\nobreak

SS was supported in part by JSPS Grant-in-Aid for Scientific Research (C) \#17K05421,  
and is grateful to Institut de Physique Th{\' e}orique, CEA-Saclay for their hospitality. 
\bigbreak

\end{document}